\documentclass[reprint,noshowpacs,noshowkeys,prd,balancelastpage,onecolumn]{revtex4}
\usepackage{amsfonts}
\usepackage{mdframed}
\usepackage{amssymb}
\usepackage{amsmath}
\usepackage{graphicx}
\usepackage{float}
\usepackage[font={footnotesize,it}]{caption}
\usepackage[utf8]{inputenc}
\usepackage{natbib}
\usepackage{tikz}
\usepackage{tikz-3dplot}
\usepackage[colorlinks=true,
            linkcolor=purple,
            urlcolor=purple,
            citecolor=blue]{hyperref}

\begin{document}

\title{Discontinuity Problem in the Linear Stability Analysis of Thin-Shell
Wormholes}
\author{S. Danial Forghani}
\email{danial.forghani@emu.edu.tr}
\author{S. Habib Mazharimousavi}
\email{habib.mazhari@emu.edu.tr}
\author{Mustafa Halilsoy}
\email{mustafa.halilsoy@emu.edu.tr}
\affiliation{Department of Physics, Faculty of Arts and Sciences, Eastern Mediterranean
University, Famagusta, North Cyprus via Mersin 10, Turkey}

\begin{abstract}
We investigate the infinite discontinuity points of stability diagram in
thin-shell wormholes. The square of the speed of sound $\beta _{0}^{2}$,\
which is expressed in terms of pressure and energy density at equilibrium on
the throat, arises with a divergent amplitude. As this is physically
non-acceptable, \ we revise the equation of state, such that by fine-tuning
of the pressure at static equilibrium, which is at our disposal, eliminates
such a singularity. The efficacy of the method is shown in Schwarzschild,
extremal Reissner-Nordstr\"{o}m, and dilaton thin-shell wormholes.
\end{abstract}

\date{\today }
\maketitle

\section{Introduction}

The concept of thin-shell wormhole (TSW) was introduced by Visser in 1989 
\cite{Visser1,Visser3}\ in the hope of keeping the idea of wormholes alive
by confining the exotic matter to a thin shell, called the throat of the
TSW. Exotic matter, which inevitably emerges in the theories of wormholes,
is an unwanted type of matter that violates the known energy conditions such
as the weak energy condition (WEC) \cite{Azreg-Ainou1}. Pre-Visser's
theories had the exotic matter distributed on certain parts of the
spacetime, if not all over it. However, Visser's so called cut-and-paste
procedure allows us to confine such a notorious matter on a very limited
part of the space, the TSW itself. Moreover, the cut-and-paste procedure has
the advantage that can be applied to a vast variety of spacetimes \cite%
{Garcia1,Varela1,Richarte1,Sharif1,Sharif2,Montelongo1,Dias1,Eiroa1,Thibeault1,Lobo1,Eiroa2,Eiroa3,Ishak1}%
, while before Visser only some certain spacetimes had the structure of a
wormhole \cite{Visser2}. It is also worth mentioning that while TSWs are
categorized as traversable wormholes, not all the wormholes are considered
to be traversable \cite{Morris1}.

Recently we have systematically analyzed the asymmetric thin-shell wormholes
(ATSWs) with different spacetimes on the two sides of the throat \cite%
{Forghani1,Forghani3}. The role of the asymmetry in the stability of the
TSWs, if there is any, has been scrutinized. In the present article we
address to a particular issue of stability which incorporates infinite
discontinuity in the stability diagrams. Starting from the Schwarzschild
TSW, it was observed that at a particular equilibrium radius of the shell,
i.e. $a_{0}=3m$ (for Schwarzschild mass $m$) there arises a divergence in
the asymptotes of the stability curves \cite{Poisson1}. The divergence
radius takes place at a finite radius which lies outside the event and
inside the cosmological horizon (if there is any), excluding the divergences
at the center and at infinity. What gives rise to such a behavior, and is it
possible to eliminate this type of divergences?

In the barotropic equation of state (EoS), the pressure $p$ and the energy
density $\sigma $\ on the shell are related to the square of the speed of
sound $\beta ^{2}$ by $\beta ^{2}=dp/d\sigma $. This relation is expressed
by $p_{0}^{\prime }=\beta _{0}^{2}\sigma _{0}^{\prime }$, where a prime
denotes derivative with respect to the radius, all evaluated at the
equilibrium radius $a_{0}$. It turns out that the mathematical structure of
the speed of sound $\beta _{0}$ is given by a fractional expression, so that
it diverges at a finite radius whenever its denominator approaches zero,
while its numerator is nonzero. This is precisely what happens at $a_{0}=3m$
in the Schwarzschild TSW. For other TSWs also the same behavior is observed
at a certain radius of the shell. Once we identify this fact we present a
recipe to eliminate such type of divergences. That comes by considering a
more general EoS (introduced under the name \textquotedblleft variable
EoS\textquotedblright\ \cite{Garcia1,Varela1}), in which the pressure
depends explicitly also on the radius of the shell. With the new EoS, the
speed of sound relation takes the form $\beta _{0}^{2}=\left( p_{0}^{\prime
}+\gamma _{0}\right) /\sigma _{0}^{\prime }$, for $\gamma _{0}=const.$, so
that the choice $p_{0}^{\prime }+\gamma _{0}=0$ will eliminate in the limit,
the singularity for the stability diagram.

We must add that, Varela in \cite{Varela1} has pointed out the solution for
a symmetric Schwarzschild TSW without addressing the cause for such anomaly,
a general removal for other TSWs, or considering asymmetrical cases. Beside
the Schwarzschild ATSW, we consider extremal Reissner-Nordstr\"{o}m (ERN),
and also the dilaton TSWs as examples to show that our method for
eliminating the infinite discontinuity perfectly works. Let us add that
since our shell takes place at finite radius, the remaining infinite
divergences in the speed of sound at $a_{0}=r_{e}$\ (event horizon) and $%
a_{0}=\infty $\ are of not physical significance.

In section $II$ we briefly explain how a TSW can be constructed by gluing
two (generally not identical) spacetimes at a common hypersurface. Section $%
III$ is devoted to the infinite discontinuity emerging in the stability
diagram of TSWs. Therein, we clarify the subject by discussing examples from
the Schwarzschild ATSW, ERN ATSW, and dilaton TSW. In section $IV$ we
explain how replacing the barotropic EoS with the variable EoS contributes
to the infinite discontinuity removal. Finally, we conclude the paper in
section $V$. All over the article, we follow the unit convention $c=8\pi
G=4\pi \varepsilon _{0}=1$, where $c$ is the speed of light, $G$\ is the
gravitational constant, and $\varepsilon _{0}$ is the permittivity of free
space\ in $3+1$ dimensions.

\section{Constructing a TSW}

To construct a TSW by Visser's method in the spherical coordinates, consider
two distinct Lorentzian spacetimes denoted by $\left( \Sigma ,g\right) ^{\pm
}$. Out of each spacetime, a subset is cut such that no singularities or
event horizons of any sort are included, i.e. $\left( \Upsilon ,g\right)
^{\pm }\subset \left( \Sigma ,g\right) ^{\pm }$ and $\left( \Upsilon
,g\right) ^{\pm }=\{x_{\pm }^{\mu }|r_{\pm }\geq a\left( \tau \right)
>r_{e}\}$, where $r_{e}$ is any existed event horizon, and $\tau $ is the
proper time on the shell $r_{\pm }=a$. Then, by pasting these two cuts at
their common timelike hypersurface $\partial \Upsilon $, such that $\partial
\Upsilon \subset $ $\left( \Upsilon ,g\right) ^{\pm }$, one creates a
complete Riemannian spacetime which provides a passage from one spacetime to
the other. The hypersurface $\partial \Upsilon $ is indeed the throat of the
TSW and contains the exotic matter. Note that, the coordinates of the two
sides of the throat $x_{\pm }^{\mu }$, and more generally, the very nature
of the two spacetimes does not necessarily need to be the same. Although
most of the authors have been tending to consider same spacetimes as the
side manifolds, recently such a mirror symmetry was broken in some studies 
\cite{Forghani1,Forghani3,Forghani2} to introduce ATSWs.

Suppose that the line element of the bulks are given by the static general
spherically symmetric metrics%
\begin{equation}
ds_{\pm }^{2}=g_{\mu \nu }^{\pm }dx_{\pm }^{\mu }dx_{\pm }^{\nu }=-f_{\pm
}\left( r_{\pm }\right) dt_{\pm }^{2}+f_{\pm }^{-1}\left( r_{\pm }\right)
dr_{\pm }^{2}+h_{\pm }\left( r_{\pm }\right) d\Omega _{\pm }^{2},
\end{equation}%
where $f_{\pm }\left( r\right) $\ and $h_{\pm }\left( r\right) $\ are
positive functions of the radial coordinates $r_{\pm }$, and $d\Omega _{\pm
}^{2}$ are unit $2$-spheres' line elements. The line element on the
hypersurface $\partial \Upsilon $\ (the throat) is given by%
\begin{equation}
ds_{\partial \Upsilon }^{2}=q_{ij}^{\pm }d\xi ^{i}d\xi ^{j},
\end{equation}%
where $\xi ^{i}$ are the local coordinates on the shell and $q_{ij}^{\pm }=%
\frac{\partial x_{\pm }^{\mu }}{\partial \xi ^{i}}\frac{\partial x_{\pm
}^{\nu }}{\partial \xi ^{j}}g_{\mu \nu }^{\pm }$ are the localized metric of 
$\partial \Upsilon $. The unit spacelike normals to the surface are also
given by $n_{\mu }^{\pm }\frac{\partial x_{\pm }^{\mu }}{\partial \xi ^{i}}%
=0 $, provided $n_{\mu }^{\pm }n_{\pm }^{\mu }=1$. To count for the
uniqueness of the TSW, $q_{ij}^{-}=q_{ij}^{+}$ must hold on the throat. In
general relativity, this is called the first of the Israel-Darmois junction
conditions \cite{Israel1}. More particularly, this condition admits that we
have $h_{+}\left( a\right) =h_{-}\left( a\right) $ at the location of the
throat. There also exists a second junction condition. This second one
imposes a discontinuity on the extrinsic curvature tensor components, given
by

\begin{equation}
K_{ij}^{\pm }=-n_{\lambda }^{\pm }\left( \frac{\partial ^{2}x_{\pm
}^{\lambda }}{\partial \xi ^{i}\partial \xi ^{j}}+\Gamma _{\alpha \beta
}^{\lambda \pm }\frac{\partial x_{\pm }^{\alpha }}{\partial \xi ^{i}}\frac{%
\partial x_{\pm }^{\beta }}{\partial \xi ^{j}}\right) ,
\end{equation}%
where $\Gamma _{\alpha \beta }^{\lambda \pm }$ are the Christoffel symbols
of the bulk spacetimes, compatible with $g_{\alpha \beta }^{\pm }$. By
introducing $S_{j}^{i}=diag(-\sigma ,p,p)$ as the stress-energy tensor of
the perfect fluid on the throat, with $\sigma $\ and $p$\ being the surface
energy density and lateral pressure, respectively, the second junction
condition admits%
\begin{equation}
\left[ K_{j}^{i}\right] _{-}^{+}-\left[ \delta _{j}^{i}K\right]
_{-}^{+}=-S_{j}^{i},
\end{equation}%
where we symbolically have $\left[ \Psi \right] _{-}^{+}=\Psi _{+}-\Psi _{-}$
for a jump in quantity $\Psi $\ passing across the throat. Going through all
the cumbersome calculations, one obtains%
\begin{equation}
\sigma =-\frac{h^{\prime }}{h}\left( \sqrt{f_{+}+\dot{a}^{2}}+\sqrt{f_{-}+%
\dot{a}^{2}}\right)
\end{equation}%
and%
\begin{equation}
p=\frac{\sqrt{f_{+}+\dot{a}^{2}}}{2}\left( \frac{2\ddot{a}+f_{+}^{\prime }}{%
f_{+}+\dot{a}^{2}}+\frac{h^{\prime }}{h}\right) +\frac{\sqrt{f_{-}+\dot{a}%
^{2}}}{2}\left( \frac{2\ddot{a}+f_{-}^{\prime }}{f_{-}+\dot{a}^{2}}+\frac{%
h^{\prime }}{h}\right) ,
\end{equation}%
where $h=h_{+}=h_{-}$,\ due to the first junction condition. Herein, an
overdot and a prime stand for a total derivative with respect to the proper
time on the throat $\tau $ and the corresponding radial coordinates $r_{\pm
} $,\ respectively. Note that all the functions are evaluated at the
location of the throat $r_{\pm }=a$. The conservation of energy, is
identified as \cite{Eiroa1}%
\begin{equation}
\sigma ^{\prime }+\frac{h^{\prime }}{h}\left( \sigma +p\right) +\frac{%
h^{\prime 2}-2hh^{\prime \prime }}{2hh^{\prime }}\sigma =0.
\end{equation}%
This latter equation is accompanied by an EoS to make possible the so-called
linear stability analysis. In this popular method, developed by Poisson and
Visser \cite{Poisson1}, Eq. (5) is recast into the form%
\begin{equation}
\dot{a}^{2}+V\left( a\right) =0,
\end{equation}%
to resemble the equation of conservation of mechanical energy with a kinetic
term $\dot{a}^{2}$ and an effective potential term%
\begin{equation}
V\left( a\right) =-\left( \frac{h\sigma }{2h^{\prime }}\right) ^{2}-\left[ 
\frac{h^{\prime }\left( f_{+}-f_{-}\right) }{2h\sigma }\right] ^{2}+\frac{%
f_{+}+f_{-}}{2}.
\end{equation}%
The potential is then being Taylor expanded about a hypothetical static
equilibrium radius $a_{0}>r_{e}$ to a quadratic term as%
\begin{equation}
V\left( a\right) =V\left( a_{0}\right) +V^{\prime }\left( a_{0}\right)
\left( a-a_{0}\right) +\frac{1}{2}V^{\prime \prime }\left( a_{0}\right)
\left( a-a_{0}\right) ^{2}+\mathcal{O}^{3}\left( a\right) .
\end{equation}%
The first two terms on the right-hand-side are zero due to the static
version of Eq. (8) and the assumption of $a_{0}$ being the equilibrium
radius, respectively. Therefore, in the very vicinity of $a_{0}$\ the
effective potential $V\left( a\right) $ is approximated by the first
non-zero term on the right-hand-side, i.e. the third term which is
proportional to $V^{\prime \prime }\left( a_{0}\right) $. If $V^{\prime
\prime }\left( a_{0}\right) >0$ ($V^{\prime \prime }\left( a_{0}\right) <0$%
), the state of the ATSW is said to be mechanically stable (unstable). It is
actually along the way measuring $V^{\prime \prime }\left( a_{0}\right) $,
that the EoS enters the calculations. An EoS is an equation that relates the
energy density $\sigma $\ and the pressure $p$ of the throat. Of the EoSs\
which have been appeared in the literature, one may enumerates the
barotropic EoS \cite{Poisson1}, the EoS of a (generalized)\ Chaplygin gas 
\cite{Bento1}, polytropic gas EoS \cite{Eid1}, phantomlike EoS\ and the
variable EoS \cite{Varela1}. Nonetheless, the barotropic EoS, mathematically
given by $p=p\left( \sigma \right) $, due to its simple, still realistic
nature, provides a useful model for the fluid's behavior on the throat of
the TSW. Therefore, following \cite{Poisson1}, here we consider the
barotropic EoS with $\beta ^{2}\equiv dp/d\sigma $,\ which implies that 
\begin{equation}
p_{0}^{\prime }=\beta _{0}^{2}\sigma _{0}^{\prime }
\end{equation}%
holds on the throat, when it is at the equilibrium radius $a_{0}$ (The
sub-index \textit{zero} indicates the value of the parameter at $a_{0}$,
i.e. $\Upsilon _{0}=\Upsilon \left( a_{0}\right) $ for each physical
variable $\Upsilon \left( a_{0}\right) $). This in turn amounts to%
\begin{equation}
\sigma _{0}^{\prime \prime }=\frac{1}{4h_{0}^{2}h_{0}^{\prime }}\left\{ %
\left[ \left( 2\beta _{0}^{2}+5\right) \left( 3\sigma _{0}+2p_{0}\right) %
\right] h_{0}^{^{\prime }3}-2\left[ \left( 2\beta _{0}^{2}+9\right) \sigma
_{0}+4p_{0}\right] h_{0}h_{0}^{\prime }h_{0}^{\prime \prime
}+4h_{0}^{2}h_{0}^{\prime \prime \prime }\sigma _{0}\right\}
\end{equation}%
for the second derivative of the energy density with respect to the radial
coordinate $a$, at the equilibrium radius $a_{0}$. This expression for $%
\sigma _{0}^{\prime \prime }$ appears naturally in $V^{\prime \prime }\left(
a_{0}\right) $. It is observed easily that for spacetimes with $h\left(
r\right) =r^{2}$, which covers a large class, this expression reduces to $%
\sigma _{0}^{\prime \prime }=\frac{2}{a_{0}^{2}}\left( 2\beta
_{0}^{2}+3\right) \left( 3\sigma _{0}+2p_{0}\right) $.

Using the static versions of Eqs. (5) and (6), together with Eqs. (7) and
(12), one can calculate the expression for $V^{\prime \prime }\left(
a_{0}\right) $ by taking the second derivative of $V\left( a\right) $ in Eq.
(9). According to the linear stability analysis method \cite{Poisson1}, $%
V^{\prime \prime }\left( a_{0}\right) $ is then set equal to zero to write $%
\beta _{0}^{2}$ in terms of $a_{0}$ (and possibly other parameters such as
mass or charge). Afterwards, $\beta _{0}^{2}$ is plotted against $a_{0}$\
(or a redefinition of $a_{0}$)\ and the regions of stability (regions
wherein $V^{\prime \prime }\left( a_{0}\right) $ becomes positive) are
specified; e.g. see Fig. 4a for $\beta _{0}^{2}$\ against $a_{0}/m$\ plotted
for a usual Schwarzschild TSW, when the matter on the throat is barotropic.
The stable regions are indicated in the figure. Generally speaking, the
graph of $\beta _{0}^{2}$\ against $a_{0}$\ may exhibit some infinite
discontinuities at some specific radii, which are the main focus of this
study. In what follows we address these infinite discontinuities and the
reason of their emergence, followed by some examples for clarification.

\section{Infinite Discontinuity in the Stability Diagram}

Previously, we observed that for the barotropic EoS we have $p_{0}^{\prime
}=\beta _{0}^{2}\sigma _{0}^{\prime }$ on the throat. Consideration of this
and the static version of Eq. (7) leads to%
\begin{equation}
\beta _{0}^{2}=\frac{p_{0}^{\prime }}{-\frac{h_{0}^{\prime }}{h_{0}}\left(
\sigma _{0}+p_{0}\right) +\frac{2h_{0}h_{0}^{\prime \prime }-h_{0}^{\prime 2}%
}{2h_{0}h_{0}^{\prime }}\sigma _{0}},
\end{equation}%
which goes to infinity once the denominator goes to zero, unless $%
p_{0}^{\prime }\rightarrow 0$ faster. Therefore, the infinite discontinuity
is fundamental and cannot be removed, for instance, by changing the
coordinates. In the frequent case where $h\left( a\right) =a^{2}$\ for side
spacetimes both, the above equation reduces to the simpler form%
\begin{equation}
\beta _{0}^{2}=\frac{p_{0}^{\prime }}{-\frac{2}{a_{0}}\left( \sigma
_{0}+p_{0}\right) }.
\end{equation}%
In such a case, $\sigma _{0}+p_{0}\rightarrow 0$\ faster than $p_{0}^{\prime
}\rightarrow 0$\ will lead to an infinite discontinuity. Here we proceed
with some examples.

\paragraph{The Schwarzschild ATSW)}

It is well-known that for a symmetric Schwarzschild TSW there exists an
infinite discontinuity at $a_{0}=3m$, where $m$\ is the central mass of the
Schwarzschild spacetime \cite{Poisson1}. More generally, for a Schwarzschild
ATSW with metric functions%
\begin{equation}
\left\{ 
\begin{array}{c}
f_{\pm }=1-\frac{2m_{\pm }}{r_{\pm }} \\ 
h_{\pm }=r_{\pm }^{2}%
\end{array}%
\right. ,
\end{equation}%
we obtain%
\begin{equation}
\sigma _{0}+p_{0}=-\frac{\left[ a_{0}-3\left( 1+\epsilon \right) m\right]
\left( a_{0}-2m\right) \sqrt{a_{0}-2\left( 1+\epsilon \right) m}+\left[
a_{0}-2\left( 1+\epsilon \right) m\right] \left( a_{0}-3m\right) \sqrt{%
a_{0}-2m}}{a_{0}^{3/2}\left[ a_{0}-2\left( 1+\epsilon \right) m\right]
\left( a_{0}-2m\right) }.
\end{equation}%
where $m_{-}\equiv $\ $m$ and $m_{+}\equiv \left( 1+\epsilon \right) m$, and 
$\epsilon \in \left[ 0,\infty \right) $\ is the mass asymmetry factor. This
has a double root at%
\begin{equation}
a_{\text{ID}_{\pm }}=\frac{3m}{8}\left[ 3\left( \epsilon +2\right) \pm \sqrt{%
9\epsilon ^{2}+4\epsilon +4}\right] ,
\end{equation}%
where the sub-index \textquotedblleft ID\textquotedblright\ stands for
infinite discontinuity. However, for the admissible domain of $\epsilon $,\
the root with the minus sign falls behind the event horizon, i.e. $a_{\text{%
ID}_{-}}<r_{e}$, whereas $a_{\text{ID}_{+}}>r_{e}$ always holds. Hence, an
infinite discontinuity is expected at $a_{\text{ID}_{+}}$, which obviously
leads to$\ a_{\text{ID}}=3m$ for a symmetric Schwarzschild TSW with $%
\epsilon =0$. Fig. 1, which illustrates $m\left( \sigma _{0}+p_{0}\right) $
versus $a_{0}/m$ for the symmetric case, explains why $\lim_{a_{0}%
\rightarrow 3m^{\pm }}\beta _{0}^{2}=\mp \infty $. 
\begin{figure}[tbp]
\includegraphics[width=80mm,scale=0.6]{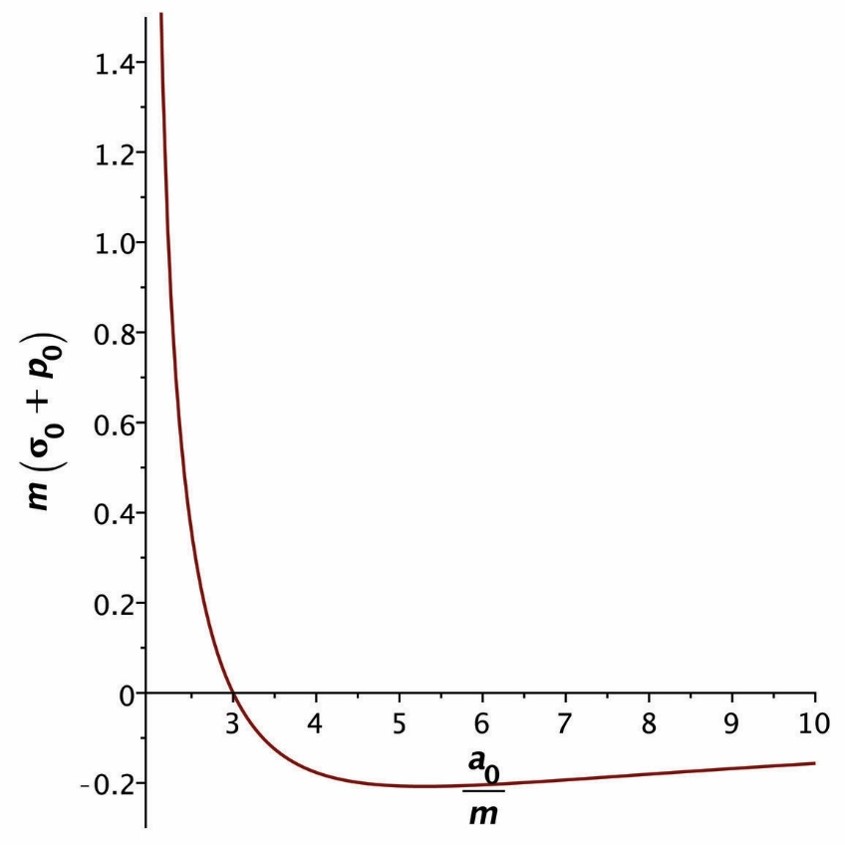}
\caption{{}The graph shows $m\left( \protect\sigma _{0}+p_{0}\right) $
versus $a_{0}/m$ for a symmetric Schwarzschild TSW. At $a_{0}=3m$, $m\left( 
\protect\sigma _{0}+p_{0}\right) =0$ which confirms the previous results.
Note that while $\protect\sigma _{0}+p_{0}$ is positive valued pre-$3m$, it
is negative post-$3m$. This explaines $\lim_{a_{0}\rightarrow 3m^{\pm }}%
\protect\beta _{0}^{2}=\mp \infty $ in the original stability diagram.}
\end{figure}

\paragraph{The Extremal Reissner-Nordstr\"{o}m TSW)}

The case of an extremal Reissner-Nordstr\"{o}m (ERN) TSW has also been
considered in the literature \cite{Eiroa2,Sharif3,Mazhari1}. Having%
\begin{equation}
\left\{ 
\begin{array}{c}
f_{\pm }=\left( 1-\frac{m_{\pm }}{r_{\pm }}\right) ^{2} \\ 
h_{\pm }=r_{\pm }^{2}%
\end{array}%
\right. ,
\end{equation}%
as the metric function of ERN, we obtain%
\begin{equation}
\sigma _{0}+p_{0}=-\frac{2\left[ a_{0}-\left( \epsilon +2\right) m\right] }{%
a_{0}^{2}}.
\end{equation}%
Accordingly, there must be an infinite discontinuity at 
\begin{equation}
a_{\text{ID}}=\left( \epsilon +2\right) m.
\end{equation}%
As shown in \cite{Eiroa2}, this also admits an infinite discontinuity at $%
a_{0}=2m$ for a symmetric ERN TSW, when $\epsilon =0$. For such a symmetric
TSW, analogous to the previous case, $\lim_{a_{0}\rightarrow 2m^{\pm }}\beta
_{0}^{2}=\mp \infty $, according to Fig. 2 plotted for $m\left( \sigma
_{0}+p_{0}\right) $ versus $a_{0}/m$. 
\begin{figure}[tbp]
\includegraphics[width=80mm,scale=0.6]{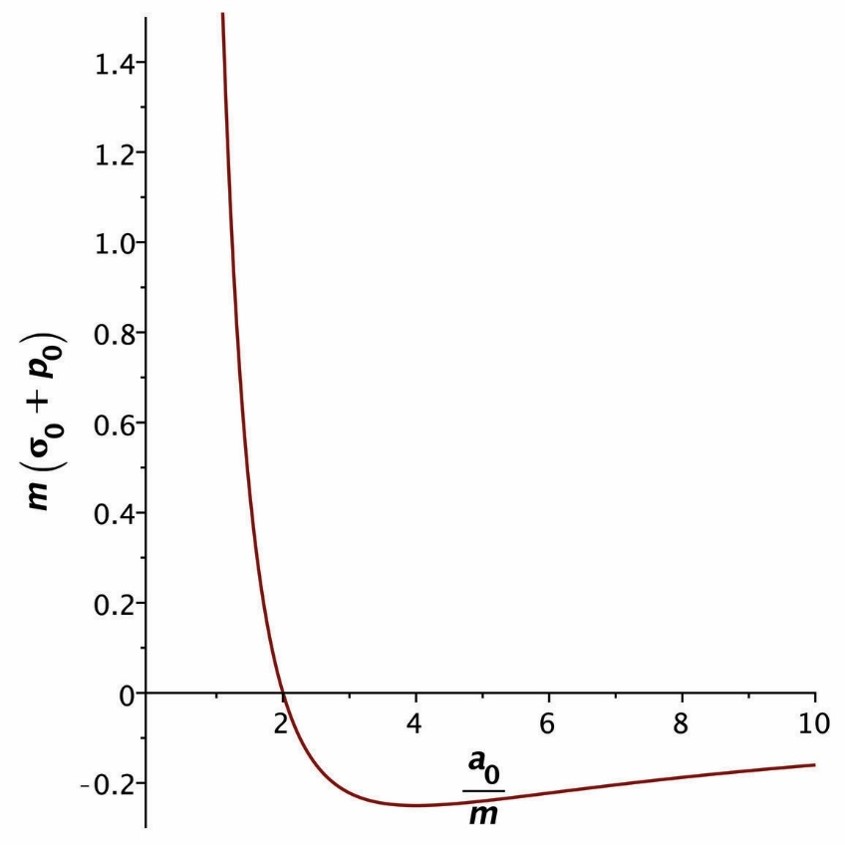}
\caption{{}{}The graph of $m\left( \protect\sigma _{0}+p_{0}\right) $
against $a_{0}/m$ for a symmetric ERN TSW. The zero of the vertical axis at $%
a_{0}=2m$ is expected according to the previous studies.}
\end{figure}

\paragraph{The Dilaton TSW)}

In \cite{Eiroa1},\ Eiroa studies a TSW constructed by two symmetric
spacetimes which are solutions of the action%
\begin{equation}
S=\int d^{4}x\sqrt{-g}\left[ -R+\left( \nabla \phi \right) ^{2}+e^{-2b\phi
}F^{2}\right] .
\end{equation}%
Herein, $g=\det \left( g_{\mu \nu }\right) $, $R$\ is the Ricci scalar, $%
\phi $\ is the scalar dilaton field, $F=F^{\mu \nu }F_{\mu \nu }$ with $%
F_{\mu \nu }$\ being the electromagnetic field, and $b\in \left[ 0,1\right] $%
\ is the coupling parameter between the dilaton and the electromagnetic
field. In Schwarzschild coordinate the spherically symmetric solution is
given by%
\begin{equation}
ds^{2}=-f\left( r\right) dt^{2}+f^{-1}\left( r\right) dr^{2}+h\left(
r\right) d\Omega ^{2}
\end{equation}%
where the metric functions are \cite{Garfinkle1,Gibbons1}%
\begin{equation}
\left\{ 
\begin{array}{c}
f\left( r\right) =\left( 1-\frac{A}{r}\right) \left( 1-\frac{B}{r}\right)
^{\left( 1-b^{2}\right) /\left( 1+b^{2}\right) } \\ 
h\left( r\right) =r^{2}\left( 1-\frac{B}{r}\right) ^{2b^{2}/\left(
1+b^{2}\right) }%
\end{array}%
\right. .
\end{equation}%
The constants $A$\ and\ $B$\ are related with the mass $m$\ and charge $q$\
of the spacetime through%
\begin{equation}
\left\{ 
\begin{array}{c}
A=m\pm \sqrt{m^{2}-\left( 1-b^{2}\right) q^{2}} \\ 
B=\left( 1+b^{2}\right) q^{2}/A%
\end{array}%
\right. .
\end{equation}%
Here we consider only the plus sign, for it is the plus sign that
corresponds to the Schwarzschild metric when $q=0$. The solutions for $b=0$\
reduce to the normal RN solutions for the Einstein-Maxwell action with a
scalar field. For $b=1$ a family of static, spherically symmetric charged
solutions in the context of low-energy string theory are recovered \cite%
{Garfinkle1}. Moreover, for $0\leq q^{2}<1+b^{2}$\ the solution is a black
hole with an event horizon at $r=A$ and an inner horizon at $r=B$. When $%
1+b^{2}\leq q^{2}\leq 1/\left( 1-b^{2}\right) $, the inner horizon grows
larger than the event horizon and the metric exhibits a naked singularity.
Also,\ the spacetime is not well-defined if $q^{2}>1/\left( 1-b^{2}\right) $%
. In what follows $a_{0}$ is considered to be greater than $A$ and $B$, as
it must be.

According to the solution in Eqs. (22) and (23), it is expected that the
root of the denominator in the expression for $\beta _{0}^{2}$\ in Eq. (13)
denotes the infinite discontinuity in the stability diagram. Having
considered, the static energy density as%
\begin{equation}
\sigma _{0}=-2\frac{h_{0}^{\prime }}{h_{0}}\sqrt{f_{0}},
\end{equation}%
and the static pressure as%
\begin{equation}
p_{0}=\left( \frac{f_{0}^{\prime }}{f_{0}}+\frac{h_{0}^{\prime }}{h_{0}}%
\right) \sqrt{f_{0}},
\end{equation}%
one calculates for the roots of the denominator of $\beta _{0}^{2}$ in Eq.
(13), to acquire $a_{\text{ID}}$.\ Due to the relatively complicated forms
of $f\left( r\right) $\ and $h\left( r\right) $\ in Eq. (23), the expression
for $a_{\text{ID}}$ is complicated and lengthy, too. For this reason, we
refrain from bringing its explicit form here. Instead, we summarize the
results in Fig. 3. Considering five different values for $b$, the subfigures
display $a_{\text{ID}}$, the event horizon ($EH$)\ associated with $r=A$,\
and the inner horizon ($IH$) associated with $r=B$, in diagrams of $a_{\text{%
ID}}/m$ against $\left\vert q\right\vert /m$. The results are in complete
agreement with the ones in \cite{Eiroa1}. For instance, for $b=0$ in Fig.
3a, there always exists an infinite discontinuity beyond horizons. As
numerical examples, this discontinuity is $a_{\text{ID}_{+}}=3m$ for $%
\left\vert q\right\vert /m=0$, and $a_{\text{ID}_{+}}\simeq 2.485m$ for $%
\left\vert q\right\vert /m=0.8$, as expected. Note that $\left\vert
q\right\vert /m>1$ is not permitted due to the restricting conditions on the
bulk spacetime mentioned above. In the case $b=1$, on the other hand, for $%
\left\vert q\right\vert /m\geq \sqrt{2}$ there is no infinite discontinuity.
Again, for the sake of comparison to the results in \cite{Eiroa1}, note
that, for example, when $\left\vert q\right\vert /m=0.8$ we obtain $a_{\text{%
ID}}\simeq 2.594m$, and $a_{\text{ID}}\simeq 2.860m$, when\ $b=0.5$, and $b=1
$, respectively. Furthermore, remark that $a_{\text{ID}}$\ in all cases is $%
3m$ when $\left\vert q\right\vert /m=0$, for the simple fact that when $q=0$%
, the metric functions in Eq. (23) reduce to the Schwarzschild metric
functions, regardless of the value of $b$.\ 
\begin{figure}[th]
\includegraphics[scale=0.3]{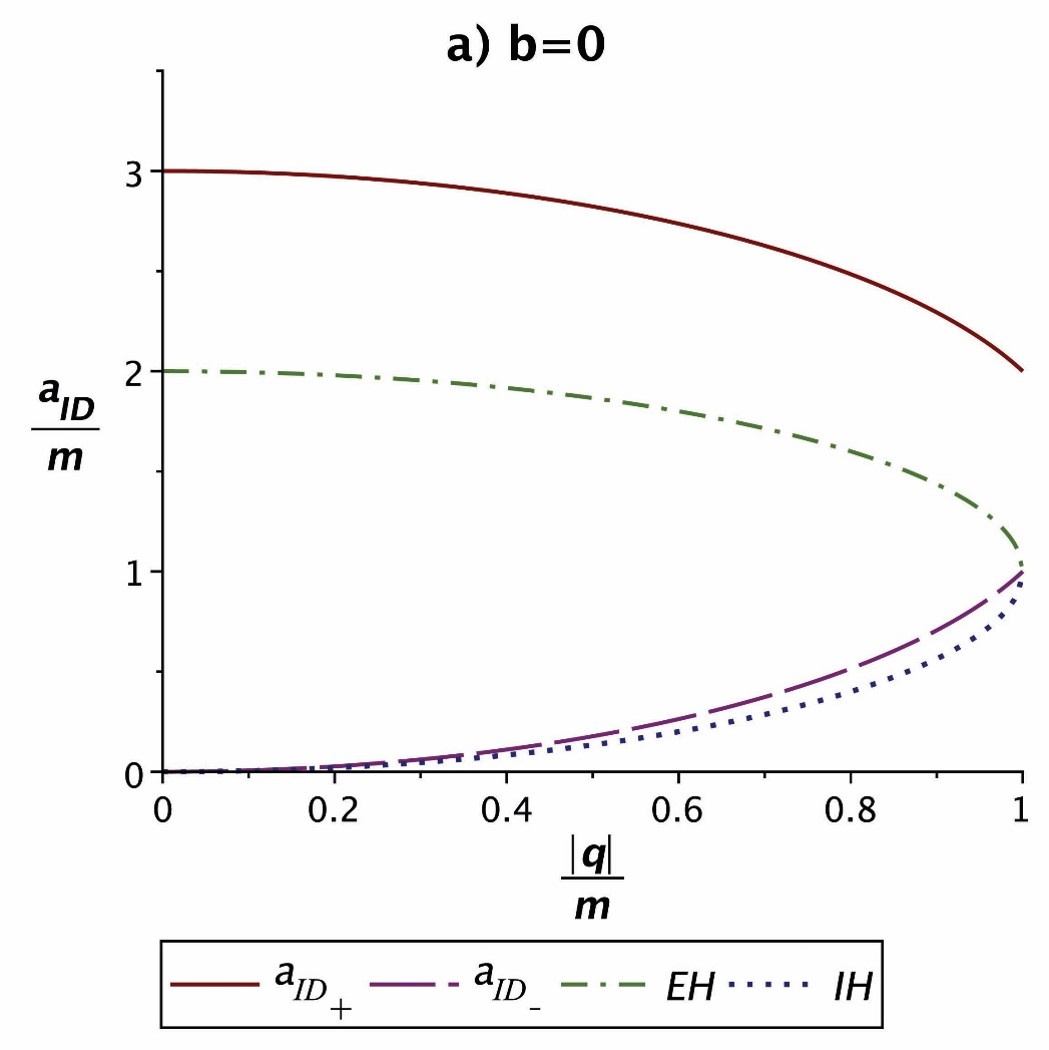} \hspace{0.1in} %
\includegraphics[scale=0.3]{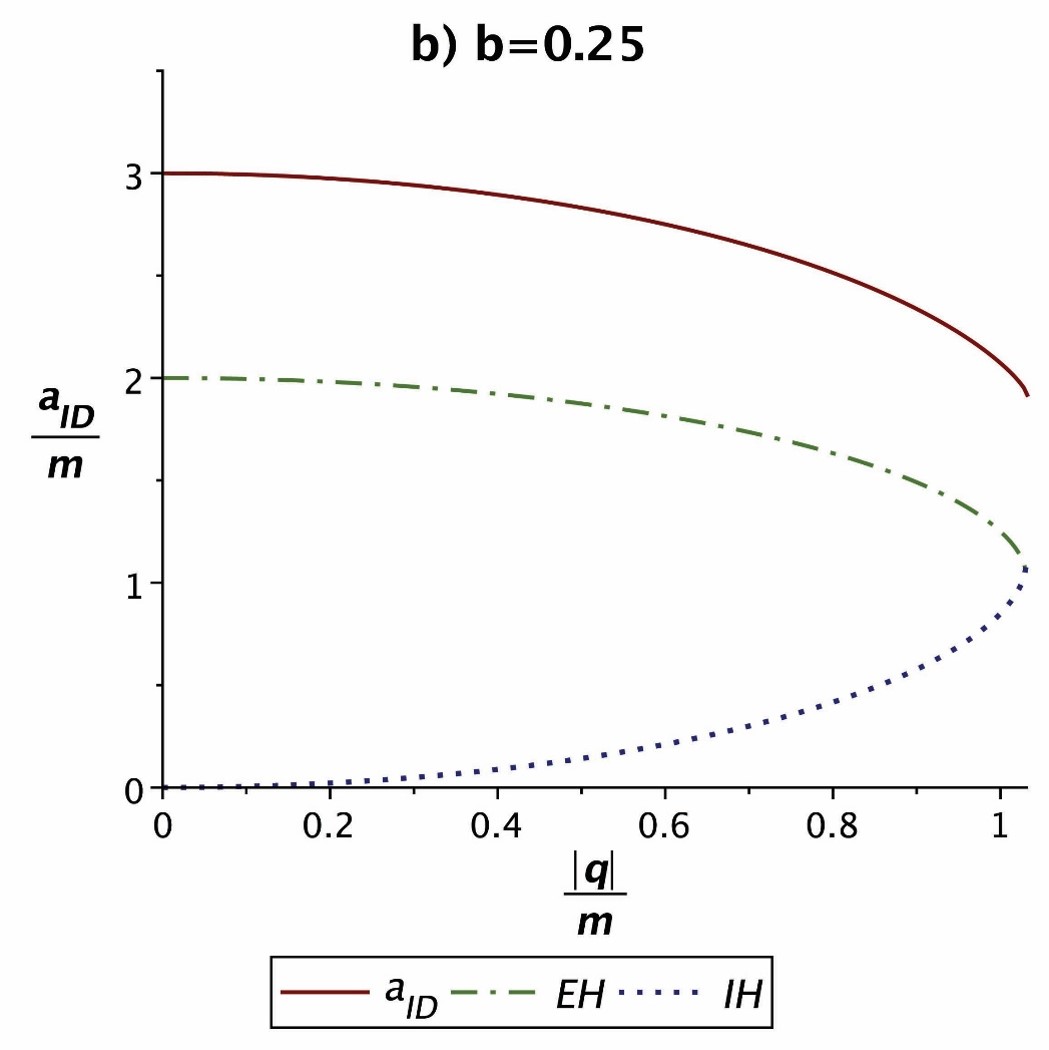} \newline
\includegraphics[scale=0.3]{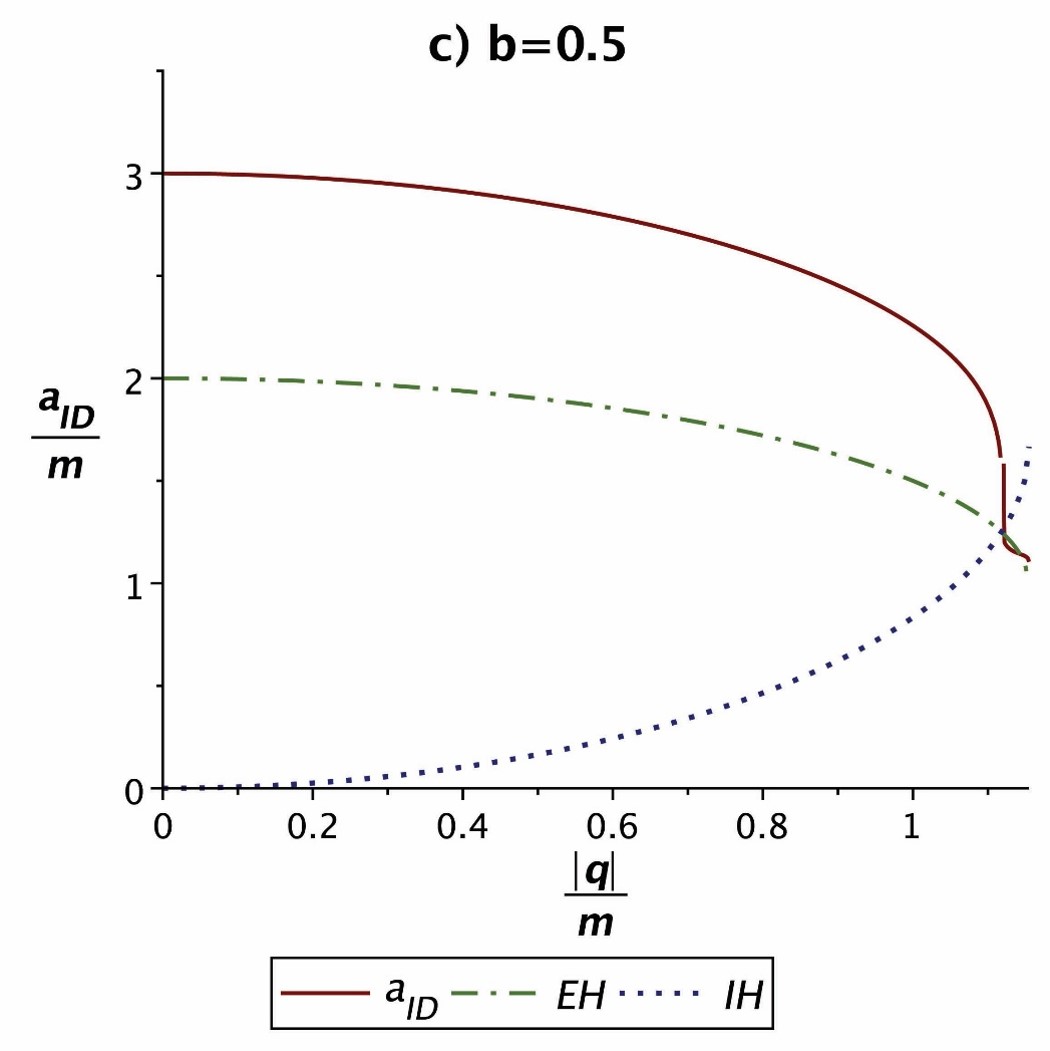} \hspace{0.1in} %
\includegraphics[scale=0.3]{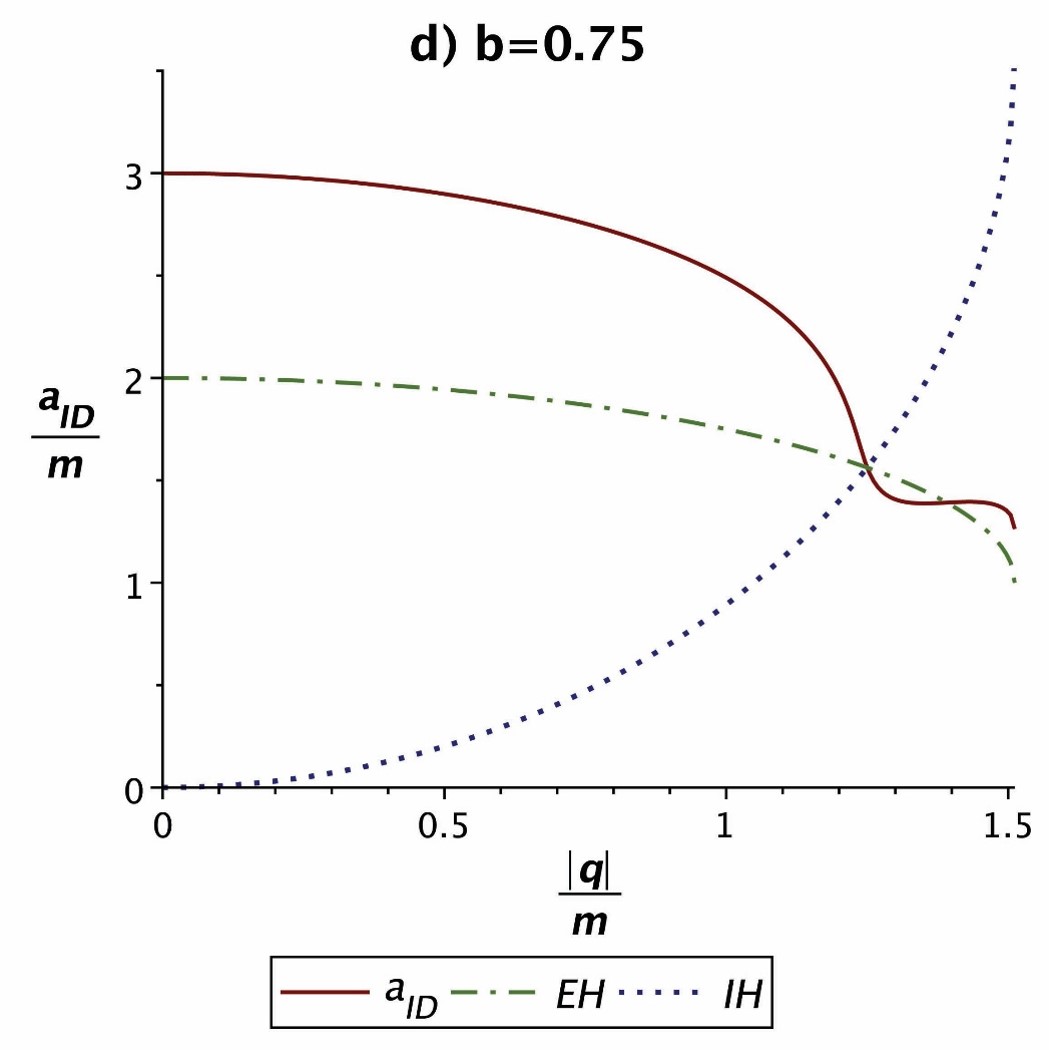} \newline
\includegraphics[scale=0.3]{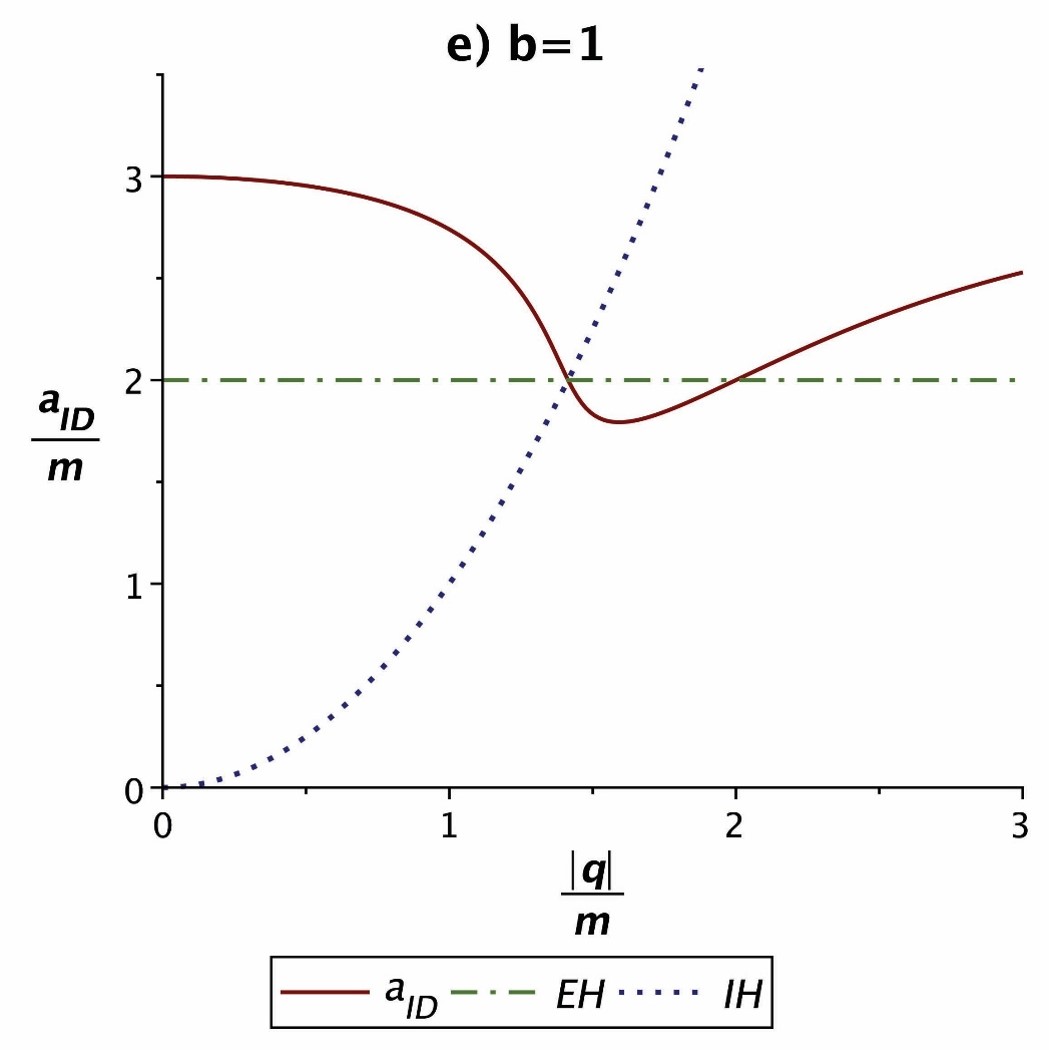} \hspace{0.1in}
\caption{The graphs show $a_{\text{ID}}/m$ against $|q|/m$ for a dilaton TSW
for different values of $b$. In the legend, $a_{\text{ID}_{+}}$ and $a_{%
\text{ID}_{-}}$ correspond to the roots of the denominator of $\protect\beta %
_{0}^{2}$, given in Eq. (30). Also, $EH$ and $IH$ correspond to the event
horizon and the inner horizon of the bulk universe.}
\end{figure}

\section{Variable EoS}

In 2015, Varela demonstrated that the infinite discontinuity of a
Schwarzschild TSW can be removed by using a rather different EoS called the
variable EoS \cite{Varela1}. Mathematically shown as $p=p\left( \sigma
,a\right) $, the variable EoS grants the pressure an explicit
radius-dependency. As a consequence, Eq. (11) will be replaced by%
\begin{equation}
p_{0}^{\prime }=\beta _{0}^{2}\sigma _{0}^{\prime }-\gamma _{0},
\end{equation}%
where now $\gamma _{0}\equiv -\left. \partial p/\partial a\right\vert
_{a_{0}}$. The mechanism of infinite discontinuity removal by the variable
EoS is simply to make null the numerator of%
\begin{equation}
\beta _{0}^{2}=\frac{p_{0}^{\prime }+\gamma _{0}}{-\frac{h_{0}^{\prime }}{%
h_{0}}\left( \sigma _{0}+p_{0}\right) +\frac{2h_{0}h_{0}^{\prime \prime
}-h_{0}^{\prime 2}}{2h_{0}h_{0}^{\prime }}\sigma _{0}}
\end{equation}%
at $a_{\text{ID}}$, where the discontinuity used to happen when $\gamma _{0}$
was zero. This means, if we set $\gamma _{0}$ such that%
\begin{equation}
\gamma _{0}=-p_{0}^{\prime },
\end{equation}%
then $\lim_{a_{0}\rightarrow a_{\text{ID}}}\beta _{0}^{2}=\frac{0}{0}$ is
indefinite, and it becomes well-defined if the numerator approaches zero, at
least, at the same rate as the denominator. As far as the unit convention
that is applied here concerns, $\beta _{0}^{2}$\ is a dimensionless quantity
($\beta _{0}$ is of type speed, with the SI dimension $\left[ LT^{-1}\right] 
$, which becomes dimensionless here, since length and time are looked at on
an equal footing in general relativity).\ Hereupon, the numerator and the
denominator of $\beta _{0}^{2}$\ have the same dimension ($L^{-2}$), and in
case the fine-tuning $\gamma _{0}=-p_{0}^{\prime }$\ is exerted, $\beta
_{0}^{2}$ can be well-defined. Note that Eq. (28) is somehow a
generalization to Eq. (13).

For the case of a symmetric Schwarzschild TSW one obtains%
\begin{equation}
\left. p_{0}^{\prime }\right\vert _{a_{0}=3m}=-\frac{2\sqrt{3}}{9m^{2}},
\end{equation}%
by taking the first derivative of Eq. (6), applying Eq. (15) and setting $%
\epsilon =0$. This means that by fine-tuning $\gamma _{0}$ to $2\sqrt{3}%
/\left( 9m^{2}\right) $ we may be free of infinite discontinuity in the
stability diagram. This is particularly shown in Fig. 4 for $\beta _{0}^{2}$
against $a_{0}/m$. As it is evident, there is no sign of the infinite
discontinuity anymore. 
\begin{figure}[tbp]
\includegraphics[scale=0.3]{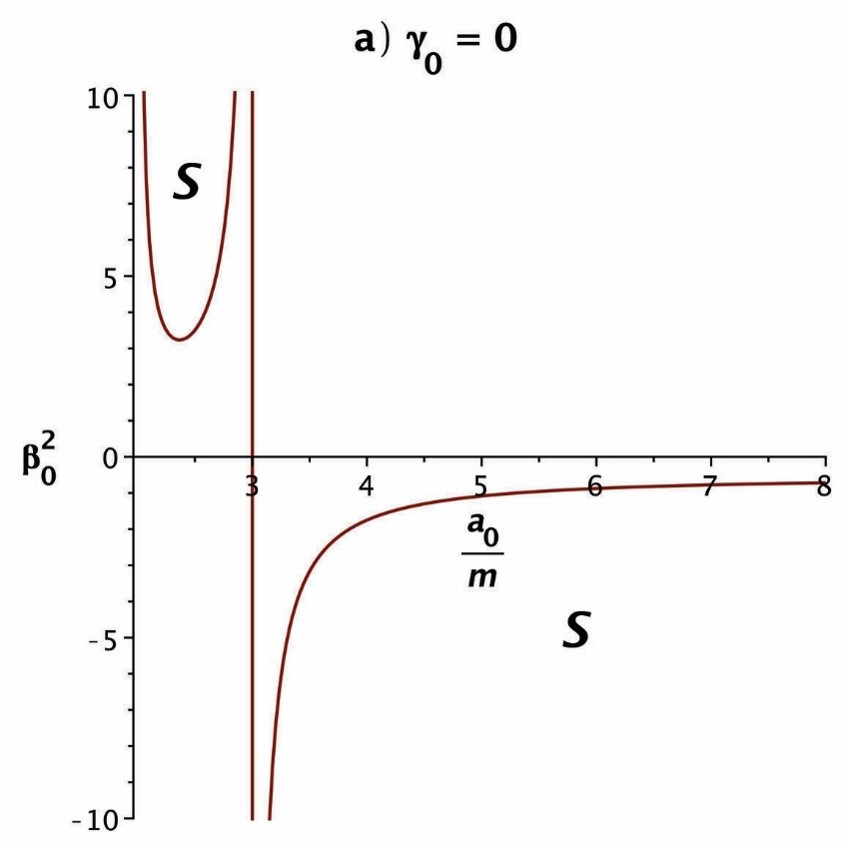} \hspace{0.1in} %
\includegraphics[scale=0.3]{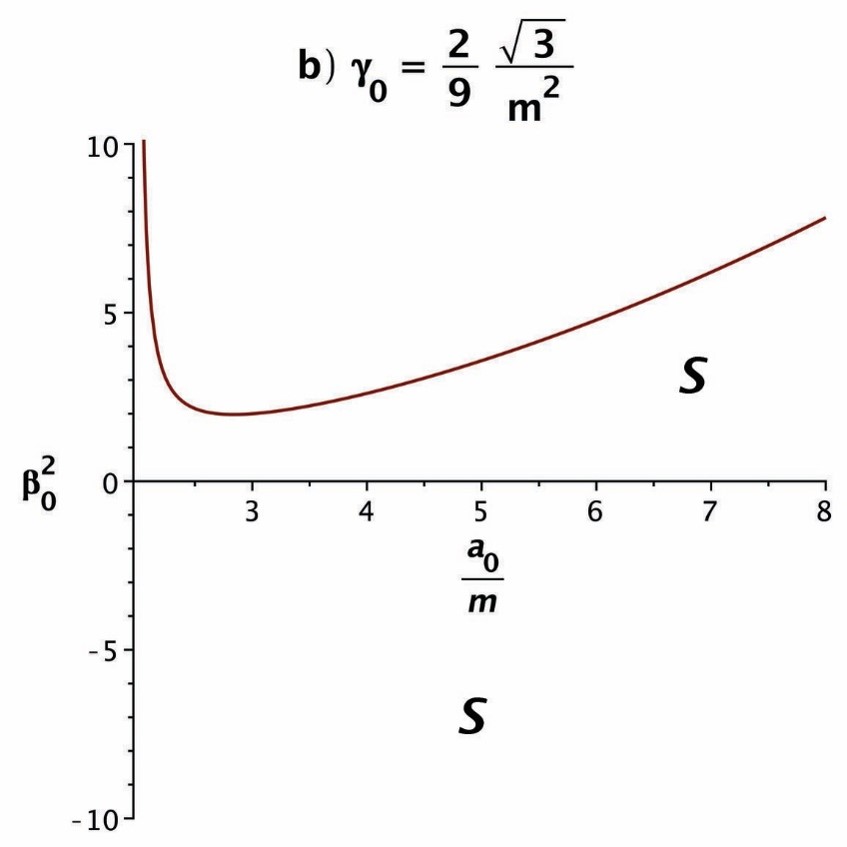} \newline
\caption{The stability diagram for a symmetric Schwarzschild TSW with $a)$
the barotropic EoS and $b)$ the variable EoS. It can be observed that the
infinite discontinuity is simply removed by virtue of the variable EoS. }
\end{figure}

Applying the variable EoS to an ERN ATSW leads to the same result. In this
case we obtain%
\begin{equation}
\left. p_{0}^{\prime }\right\vert _{a_{0}=\left( \epsilon +2\right) m}=-%
\frac{2}{\left( \epsilon +2\right) ^{2}m^{2}},
\end{equation}%
as the derivative of the angular pressure at the radius of infinite
discontinuity occurrence. Correspondingly, the choice $\gamma _{0}=2/\left(
\left( \epsilon +2\right) ^{2}m^{2}\right) $ is expected to remove the
infinite discontinuity. Fig. 5 shows how this happens for a symmetric ERN
TSW, for which $\epsilon =0$. 
\begin{figure}[tbp]
\includegraphics[scale=0.3]{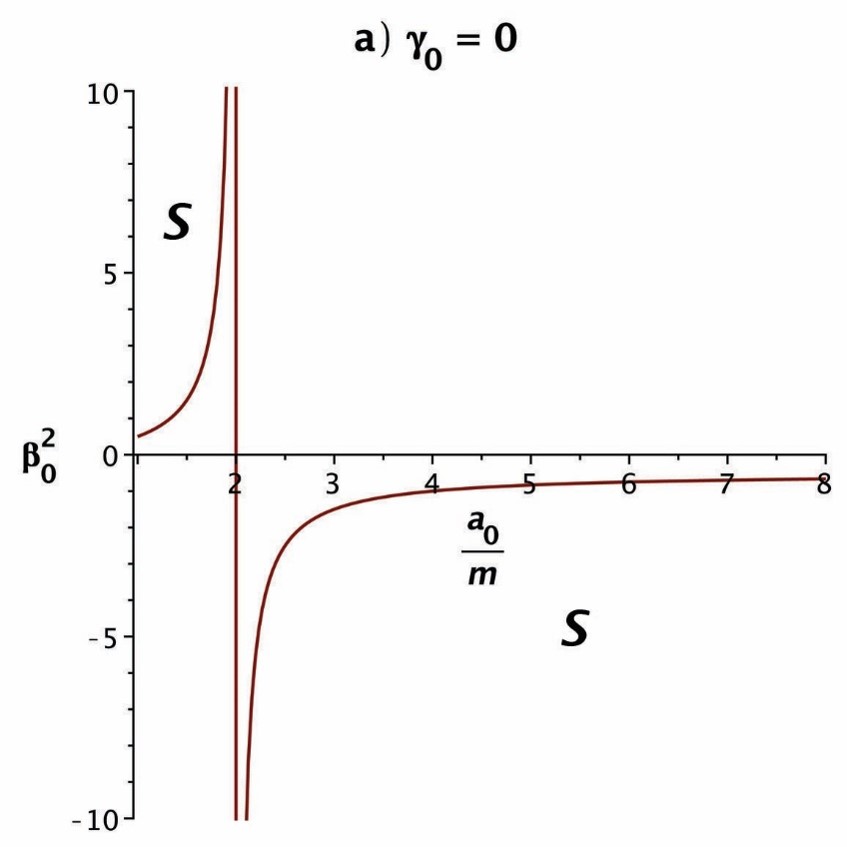} \hspace{0.1in} %
\includegraphics[scale=0.3]{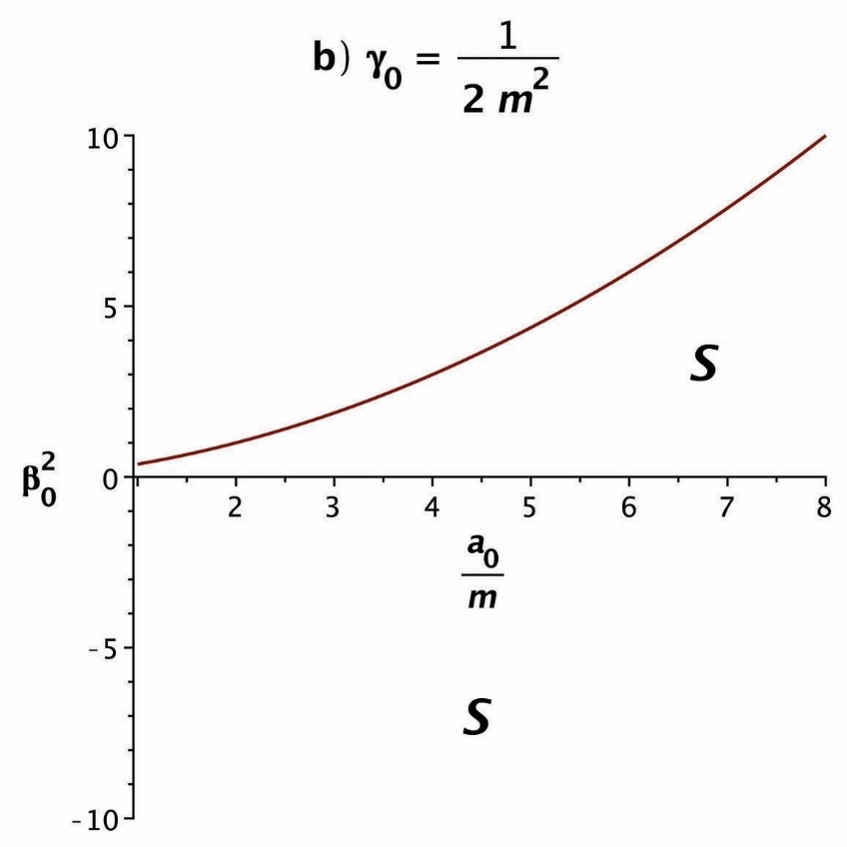} \newline
\caption{The stability diagram for a symmetric ERN TSW with $a)$ the
barotropic EoS and $b)$ the variable EoS. As expected, the infinite
discontinuity is removed due to the fine-tunning of the variable EoS. }
\end{figure}

In the end, we turn our attention to the dilaton TSW. Here we will have a
closer look at three cases for which $b=0$, $b=0.5$, and $b=1$. The former
is selected for it defines the RN spacetime, and the latter is selected for
its importance in string theory. The choice $b=0.5$\ is rather random, as a
middling value in the $b$-spectrum.\ Also, for all three cases, without loss
of generality, we have randomly chosen $\left\vert q\right\vert /m=0.5$. Our
numerical analysis shows that for the three case we have%
\begin{equation}
\left\{ 
\begin{array}{c}
\left. p_{0}^{\prime }\right\vert _{a_{0}=a_{\text{ID}}}\simeq
-0.4139864432/m^{2}\text{ \ \ when \ \ }b=0 \\ 
\left. p_{0}^{\prime }\right\vert _{a_{0}=a_{\text{ID}}}\simeq
-0.4146553639/m^{2}\text{ \ \ when \ \ }b=0.5 \\ 
\left. p_{0}^{\prime }\right\vert _{a_{0}=a_{\text{ID}}}\simeq
-0.4162095590/m^{2}\text{ \ \ when \ \ }b=1%
\end{array}%
\right. ,
\end{equation}%
which denotes that if we fine-tune $\gamma _{0}$ such that $\left. \gamma
_{0}\right\vert _{b=0}=0.4139864432/m^{2}$, $\left. \gamma _{0}\right\vert
_{b=0.5}=0.4146553639/m^{2}$, and $\left. \gamma _{0}\right\vert
_{b=1}=0.4162095590/m^{2}$, the existed discontinuities must be removed.
This can be seen clearly in Fig. 6, where the related mechanical stability
diagrams are plotted for the three cases, once when $\gamma _{0}=0$
(barotropic EoS), and once when it is fine-tuned to remove the
discontinuity. In all the subfigures, the horizontal axis starts at $A/m$,\
and the stable regions are marked with an \textquotedblleft \textsf{S}%
\textquotedblright . Note that a similar analysis can be applied to other
admissible values of $b$\ and/or $\left\vert q\right\vert /m$. 
\begin{figure}[tbp]
\includegraphics[scale=0.3]{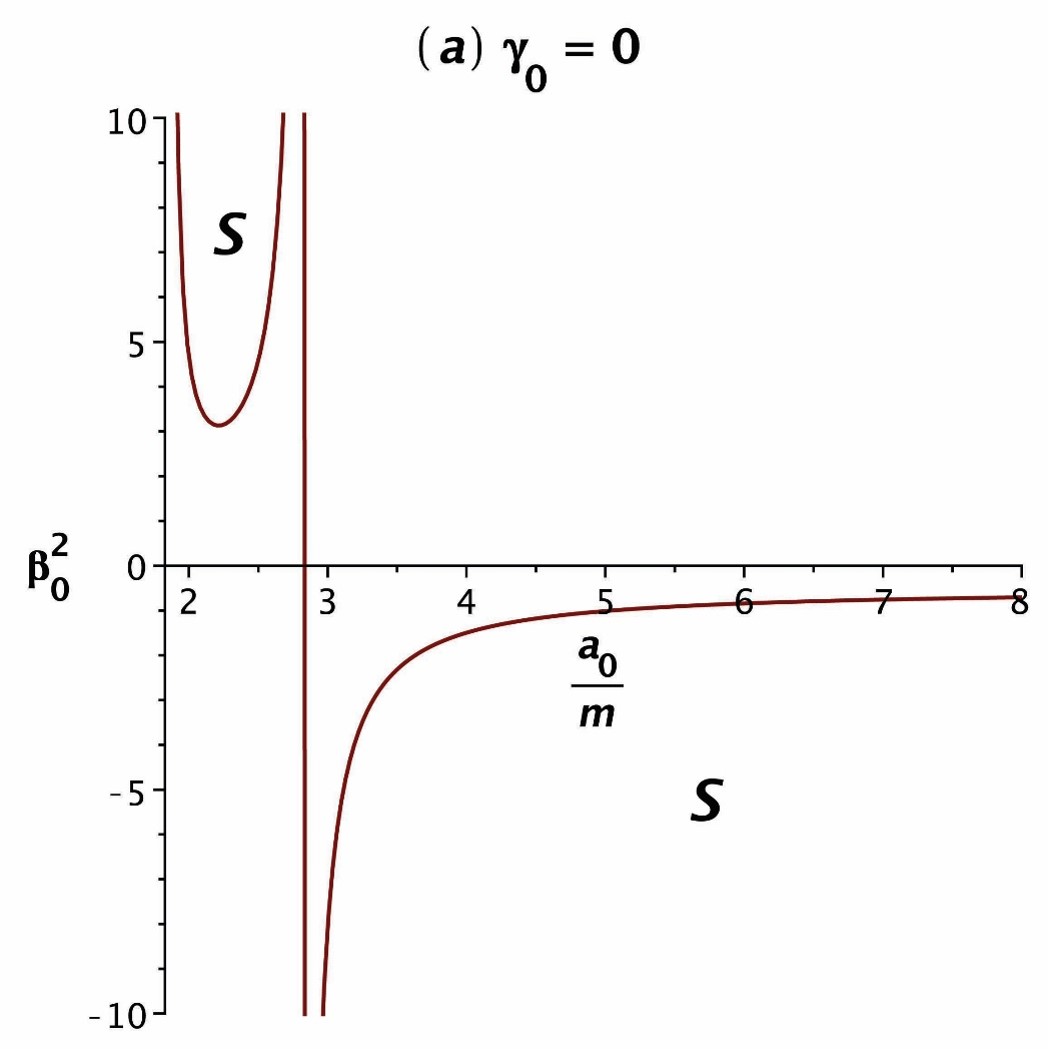} \hspace{0.1in} %
\includegraphics[scale=0.3]{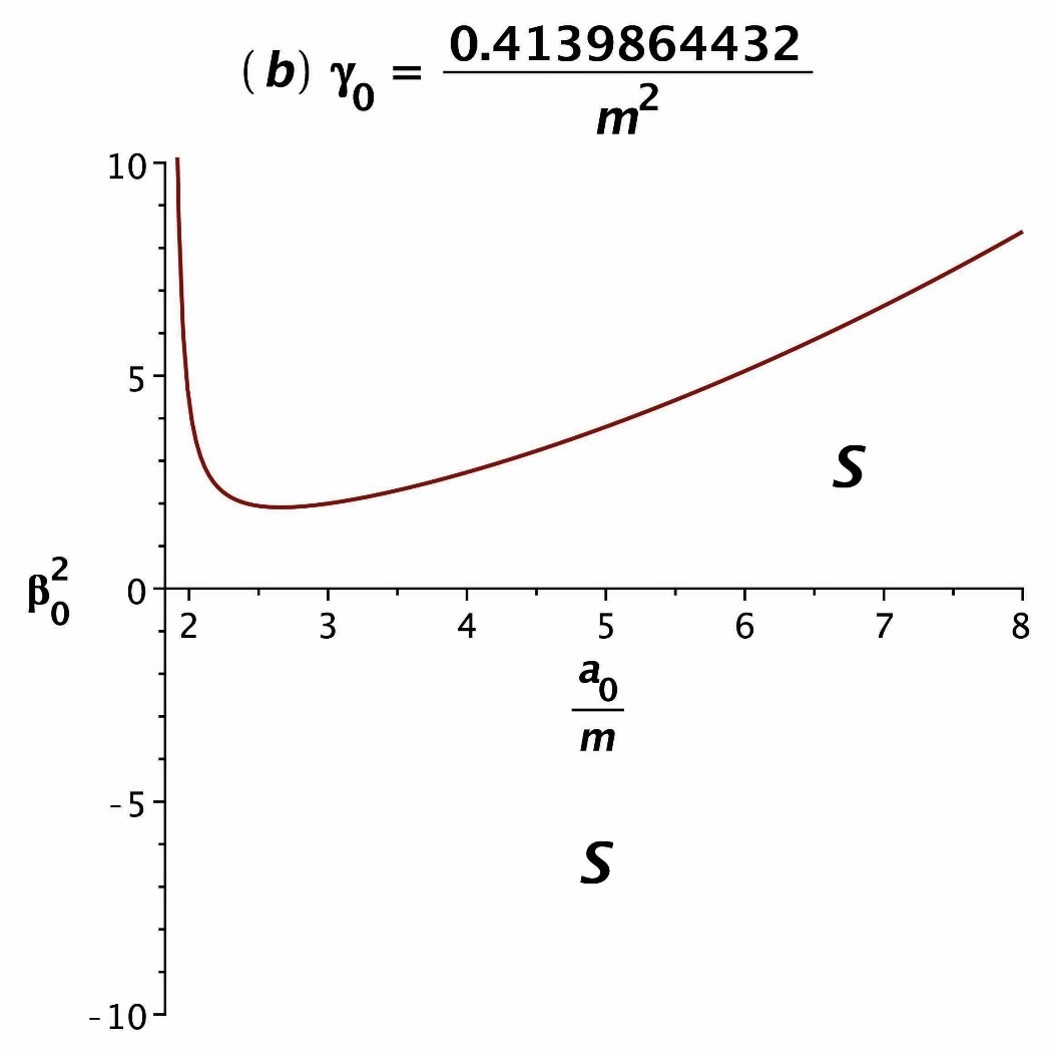} \newline
\includegraphics[scale=0.3]{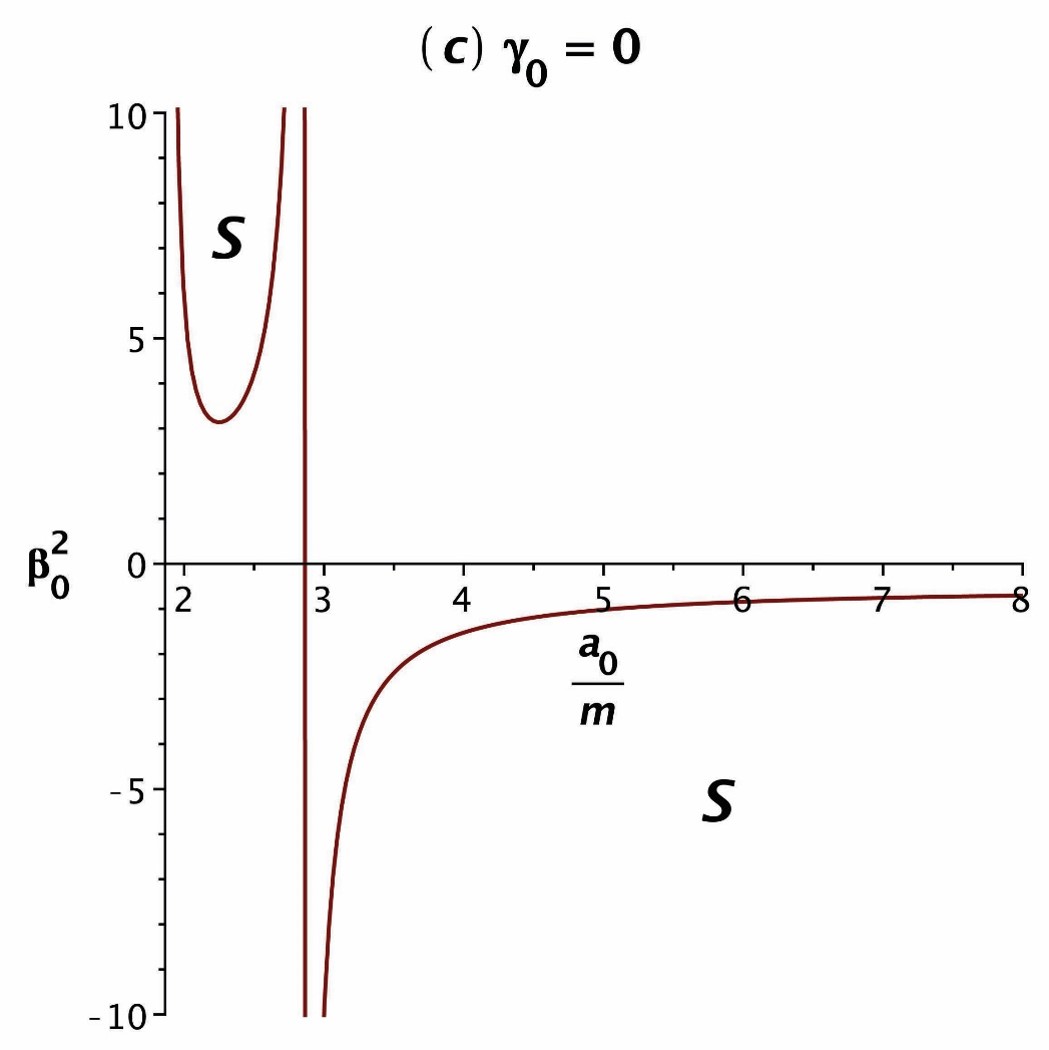} \hspace{0.1in} %
\includegraphics[scale=0.3]{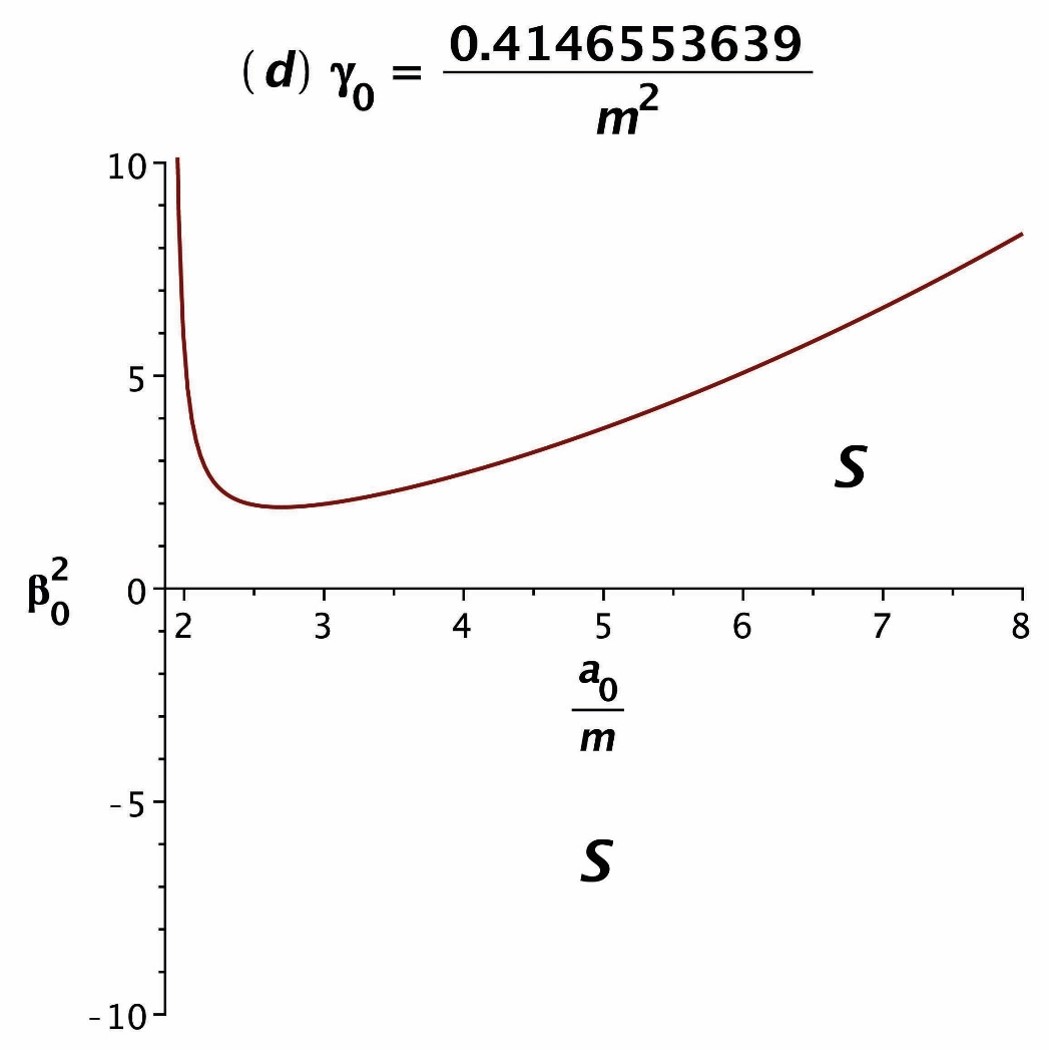} \newline
\includegraphics[scale=0.3]{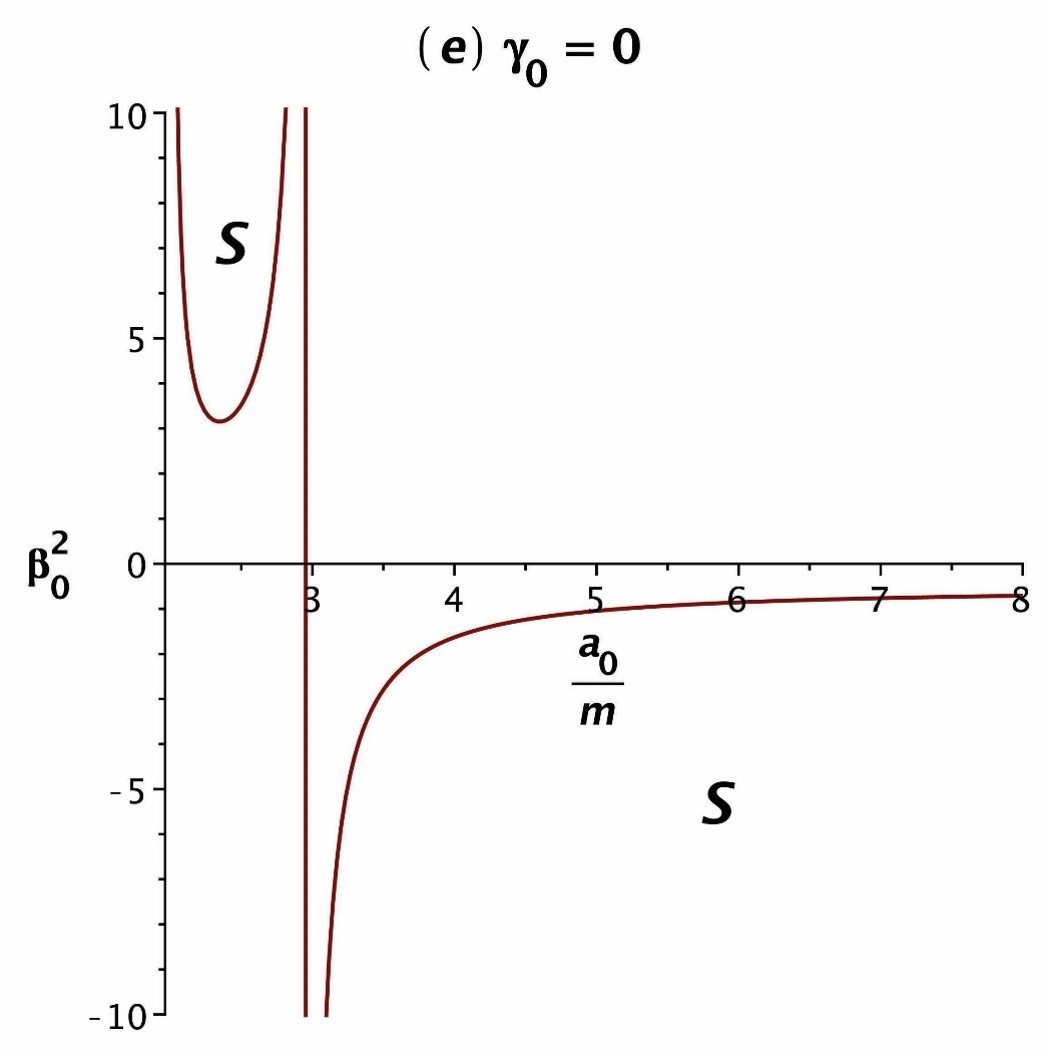} \hspace{0.1in} %
\includegraphics[scale=0.3]{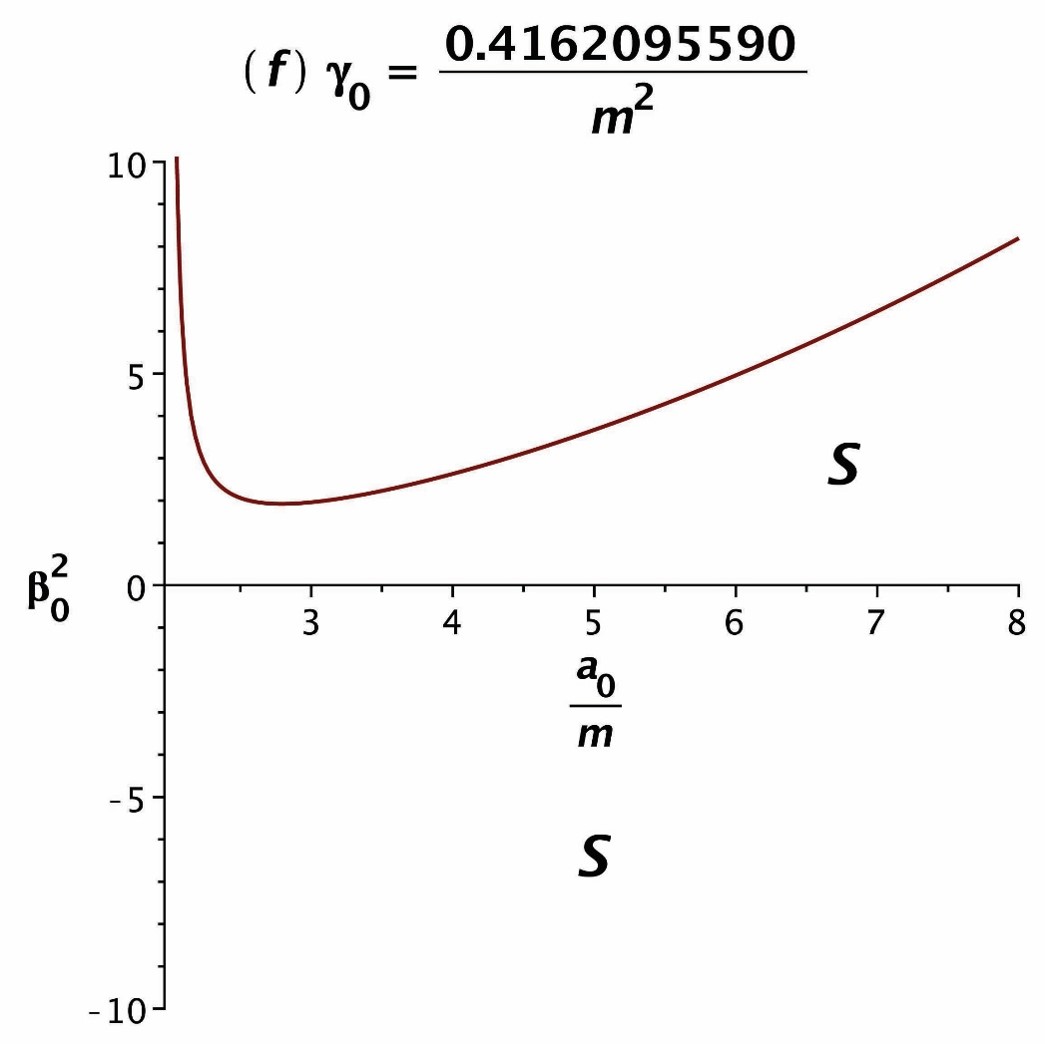} \newline
\caption{The stability diagram for a symmetric dilaton TSW for $a)$ $b=0$
with a barotropic EoS, $b)$ $b=0$ with a fine-tuned variable EoS, $c)$ $b=0.5
$ with a barotropic EoS, $d)$ $b=0.5$ with a fine-tuned variable EoS, $e)$ $%
b=1$ with a barotropic EoS, and $f)$ $b=1$ with a fine-tuned variable EoS.
The value of $\left\vert q\right\vert /m$ is set half for all the cases. The
fine-tuned values of $\protect\gamma _{0}$ are given at the top of each
diagram. The regions with \textquotedblleft \textsf{S}\textquotedblright\
are where the TSW is mechanically stable.}
\end{figure}

\section{Conclusion}

The emergence of infinitely branching discontinuity, resembling a phase
transition, seemed peculiar enough to attract attention since the stability
analyses for TSWs were incepted. The prototype example was the Schwarzschild
TSW which had such a discontinuity at the stability radius $a_{0}=3m$, as
pointed out by Poisson and Visser \cite{Poisson1}. In analogy, other TSWs
also exhibited similar behavior. We identified the cause of such type of
discontinuities: they arise from the vanishing of $-\left( h_{0}^{\prime
}/h_{0}\right) \left( \sigma _{0}+p_{0}\right) +\left( 2h_{0}h_{0}^{\prime
\prime }-h_{0}^{\prime 2}\right) \sigma _{0}/2h_{0}h_{0}^{\prime }$\ at
equilibrium radius (Eq. (13)). In consequence, the speed of sound $\beta
_{0} $, which is inversely proportional to this expression, and is expected
to be finite, naturally diverges. In an attempt to resolve such a
discontinuity in the symmetric Schwarzschild TSW, Varela employs a more
general, modified EoS, i.e. the variable EoS, to replace the barotropic one.
In this rather general EoS, the pressure $p$\ depends, beside $\sigma $,
also on the radius of the shell which creates an extra degree of freedom to
be used as an advantage. We have precisely shown that such a generalization
can be systematically applied to all the TSWs, by pointing out to the reason
of the emergence of the discontinuities. Our investigation is generalized to
all spherically symmetric spacetimes, with the generic line element in Eq.
(1), including non-asymptotically flat ones such as the dilaton TSW in
section $IIIc$. This shows that the method is applicable even to those TSWs
which are strongly coupled with the non-linear, non-asymptotically flat,
dilatonic bulk spacetimes. In section $IV$,\ the logic was illustrated by
representing three examples (see Fig. 4 for the Schwarzschild, Fig. 5 for
ERN TSWs and Fig. 6 for the dilaton TSW), where as a result, the
discontinuities in question are eliminated. It is not difficult to
anticipate that the same technique can be applied also to other TSWs,
including the ones in alternative theories. Finally, let us add that in the
present article, we used asymmetric TSW in the sense that the spacetimes on
different sides of the throat differ only parametrically. No doubt, the
spacetimes that differ in $r$-dependence also can be considered within the
range of application. %

\end{document}